\documentclass{ws-procs9x6}
\usepackage{rotating}

\begin{document}

\title{%
Open charm production in binary reactions 
within the Regge theory%
}%

\author{%
A.~Yu. ILLARIONOV \ and \ G.~I. LYKASOV
}%

\address{%
Joint Institute for Nuclear Research, \\
141980 Dubna, Moscow region, Russia
}%


\maketitle

\abstracts{%
We analyze the open charm production in the binary 
$\pi \, p \to \bar{D} \, \Lambda_c$ and $\pi(\rho) \, J/\psi$
reactions within the Regge theory including the absorption 
corrections. The calculations show that the total cross section
is about a few $\mu$b for the first processes and a few mb  
for the second reactions at energy close to the threshold.
Then it decreases with increasing energy according 
to the true Regge asymptotics.
}%

In the last decade the problem of searching for a quark--gluon plasma (QGP)
has been rising in line with the development of new experimental facilities
\cite{QM:2002}. For instance, the $J/\psi$-meson plays a key role in the
context of a phase transition to the QGP where charmonium ($c\bar{c}$)
states should no longer be formed due to color screening. However, the
suppression of the $J/\psi$ and $\psi^\prime$ mesons in the high-density 
phase of nucleus--nucleus collisions might also be attributed to inelastic
comover scattering, (see, for example, \cite{Cassing:1999,Capella:2000} and
references therein) provided that the corresponding $J/\psi$-hadron cross
sections are in the order of a few mb \cite{Haglin:2000,Lin:2001,Oh:2000}. 
Present theoretical estimates differ by more than an order of
magnitude, especially with respect to $J/\psi$-meson scattering, so that
the question of charmonium suppression is not yet settled. Moreover, the
calculation of these cross sections within the chiral Lagrangian approach
results in their constant or slowly increasing energy dependence 
\cite{Haglin:2000,Lin:2001,Oh:2000}, which contradicts the true Regge 
asymptotics predicting the decreasing one when the energy increases. The
inclusion of the meson structure and the introduction of the meson form
factors in this Lagrangian model leads to a big uncertainty for the shape
and magnitude of the $J/\psi$ breakup cross sections by mesons. 


The amplitude of the reaction in question has to satisfy the Regge
asymptotics at large $s$.
In Refs.~\cite{Boreskov:1983,Rzjanin:2001} the cross section of the reaction
$\pi \, N \to \bar{D}(\bar{D}^*) \, \Lambda_c$ was estimated within the
framework of the Quark--Gluon String Model (QGSM) developed in
Ref.~\cite{Kaidalov:1982}. The QGSM is a nonperturbative approach based on
the ideas of a topological $1/N$ expansion in QCD and on the Regge theory.

We apply such an approach to the analysis of the processes like
$\pi \, N \to \bar{D} \, \Lambda_c(\Sigma_c)$ and
$\pi(\rho) \, J/\psi \to \bar{D} D^*(\bar{D}^* D^*, \bar{D} D)$. 
The amplitude for such reactions corresponding to the planar graph with
$u$ and $\bar c$ quark exchange in the $t$ channel can be written as
(see Ref. \cite{Boreskov:1983} and \cite{Rzjanin:2001})
\begin{equation}
 \mathcal{M}(s,t) \ = \ C_I \, g_0^2 \, F(t) \,
  (s/s_0)^{\alpha_{u\bar{c}}(t) -1} \, (s/\bar{s}) \ ,
\label{def:regampl}
\end{equation}
where
 $g_0^2/4\pi = 2.7$ is determined from the width of the $\rho$-meson
 \cite{Boreskov:1983};
the isotopic factor
$C_I = \sqrt{2}$ for the $\pi^\pm \, N$,  $\pi(\rho)^\pm \, J/\psi$
reactions  and $C_I =1$ for the $\pi^0 \, N$, $\pi^0(\rho^0) \, J/\psi$
reactions;
$\alpha_{u \bar{c}}(t) = \alpha_{D^*}(t) $ is the
$D^*$ Regge trajectory, $\bar{s} = 1~\text{GeV}^2$ is a universal
dimensional factor, $s_0 = 4.0~\text{GeV}^2$ is the flavor dependent
scale factor which is determined by the mean transverse mass and the
average-momentum fraction of quarks in colliding hadrons
\cite{Boreskov:1983}, and $F(t)$ is the form factor describing the $t$
dependence of the residue.
We assume, as in Refs.~\cite{Boreskov:1983,Rzjanin:2001}, that the $D^*$
Regge trajectory is linear and therefore can be expanded over the transfer
$t$
\begin{equation}
\alpha_{D^*}(t) \ = \ \alpha_{D^*}(0) \ + \ \alpha_{D^*}^\prime(0) \, t \ ,
\end{equation}
where the intercept $\alpha_{D^*}(0) = -0.86$ and its derivative
$\alpha_{D^*}^\prime(0) = 0.5~\text{GeV}^{-2}$ are found from their
relations to the same quantities for the $J/\psi$ and $\rho$ trajectories
which are known, see Ref.~\cite{Boreskov:1983}.
The form factor $F(t)$ was presented in Ref.~\cite{Boreskov:1983} as 
\begin{equation}
 F(t) \ = \ \Gamma(1 - \alpha_{D^*}(t)) \ ,
\label{def:ft}
\end{equation}
where $\Gamma(x)$ is the Gamma--function.
Note, that in the region of negative $t$ for the reactions 
$\pi \, N \to \bar{D} \, \Lambda_c (\Sigma_c)$ $F(t)$
exhibits a fractial growth (which is faster than 
exponential) and therefore is not acceptable. For this type of reactions
we will use the conventional parameterization
\cite{Boreskov:1983,Rzjanin:2001}
\begin{equation}
 F(t) \ = \ \Gamma(1 - \alpha_{D^*}(0)) \ .
\label{def:ftap}
\end{equation}

As is shown in Ref.~\cite{Kaidalov:1994}, at intermediate energies
the absorption corrections due to the elastic and inelastic rescattering
of final hadrons produced in binary reactions can be very sizable.
They can greatly reduce the magnitude of the cross section especially
at energies close to the threshold. This is why we have to include these
effects. We estimate these absorption corrections using the standard method
of reggeon calculus and the quasieikonal approximation.  The amplitude of a
binary  reaction in the impact parameter space is represented as
\cite{Kaidalov:1994}
\begin{equation}
\mathcal{M}(s,b) \ = \ \mathcal{M}_R(s,b) \, \exp\left(-\chi(s,b)\right) \ ,
\label{def:abs}
\end{equation}
where $\mathcal{M}_R(s,b)$ is the $b$-space representation of the simple
Regge-pole exchange amplitude
\begin{equation}
\mathcal{M}_R(s,b) \ = \ \int \dfrac{d^2\mathbf{q}_\perp}{2\pi} \,
 \mathcal{M}_R(s, \mathbf{q}_\perp^2) \,
 \exp\left(i \mathbf{b}\mathbf{q}_\perp \right)
\label{def:mrb}
\end{equation}
and the amplitude $\mathcal{M}_R(s,\mathbf{q}_\perp^2)$ in our case is given by
Eq.~(\ref{def:regampl}).
The function $\chi(s,b)$ in Eq.~(\ref{def:abs}) includes the possible
elastic and inelastic rescatterings of the final charmed mesons. The
elastic $\bar{D} \, \Lambda_c$ or $\bar{D}D^*$ scattering is determined
mainly by the one-Pomeron exchange graph at $s > s_\text{thr}$,
therefore~\cite{Kaidalov:1994}
\begin{equation}
 \chi(s,b) \ = \ \dfrac{C \, \sigma^\text{tot}}{4\pi\Lambda(s)} \,
  \exp\left(-\dfrac{b^2}{2\Lambda(s)}\right) \ ,
\label{def:chin} 
\end{equation}
where $\sigma^\text{tot}$ is the total cross section of the interaction
of final $\bar{D} \, \Lambda_c$ or $D(D^*)$ mesons; $\Lambda(s)$ is the
slope of the differential cross section of $D \, \Lambda_c$ or
$D^*\bar{D}(D^*\bar{D}^*,D\bar{D})$ elastic scattering. For the
one-Pomeron exchange graph
\begin{equation}
 \Lambda_\mathcal{P}(s) \ = \ 2\alpha_\mathcal{P}^\prime(0) \, \ln(s/s_0)
\label{def:lamb}
\end{equation}
where $\alpha_\mathcal{P}^\prime(0) \simeq 0.2~\text{(GeV/\textit{c})}^{-2}$
is the slope of the Pomeron trajectory.
Returning from the $b$-representation of the scattering amplitude,
given by Eq.~(\ref{def:abs}), to the momentum space we can calculate the
${\mathcal M}(s,t)$ including the absorption corrections. 
 Finally, instead of Eq.~(\ref{def:regampl}) we have the following form
of the scattering amplitude including the absorption corrections
\begin{gather}
{\mathcal M}(s,t) \, = \, C_I \, g_0^2 \, F(0) \,
 \dfrac{\left(s/s_0\right)^{\alpha_\mathcal{D^*}(0)-1} \left(s/{\bar s}\right)}
       {\Lambda_\mathcal{D^*}(s)}
 \Bigl(\dfrac{s}{s_0}\Bigr)^{\alpha_\mathcal{D^*}^\prime(0)(q_0^2 - q_z^2)}
\nonumber \\ \times
 \int_0^{\infty} f_\mathcal{D^*}(s,b) \;
 C_A(s)^{f_\mathcal{P}(s,b)} \; j_0(bq_\perp) \; bdb \ ,
\label{def:rcoramp}
\end{gather}
where $t=q^2=q_0^2 - q_z^2 - q_\perp^2$,
$j_0(x)$ is the Bessel function of the zero order,
$C_A(s) = \exp\left(-\chi(s, 0)\right)$ and
$f_R(b) = \exp(-b^2/(2\Lambda_R(s)))$.
The ``enhancement factor'' $C \simeq 1.5$ has been found in
Ref.~\cite{Kaidalov:1974} and \cite{Kaidalov:1994} to be
in good agreement with the experiment for elastic $\pi p, Kp, {\bar p} p$
scatterings. So, in our calculations we have taken the same value for
$C$ entering the Eq.~(\ref{def:chin}). The values for the total cross 
section of the scattering of final $D$-mesons can be calculated at 
$s > s_\text{thr}$ within the one-Pomeron (secondary Reggeons) exchange
graphs and with some estimates of the $D$-meson radius from
\cite{Faessler:2002}, it is
$\sigma_{D\bar{D}^*}^\text{tot} \simeq 10 \div 12~\text{mb}$,
and we assume that
$\sigma_{D\Lambda_c}^\text{tot} \simeq 3/2 \, \sigma_{D\bar{D}^*}$.

The differential cross section for the discussed reactions is then
\begin{equation}
 \dfrac{d\sigma}{dt} \ = \
  \dfrac{1}{64\pi s \, p_{\text{c.m.}}^2}
  \sum_\text{isospin}
  \left| {\mathcal M}(s,t) \right|^2 \ ,
\label{def:dsig}
\end{equation}
where $p_\text{c.m.}$ is the initial momentum in the c.m.s.
The total cross section of the process discussed is calculated
as the integral of Eq.~(\ref{def:dsig}) over $t$. 
\begin{figure}[t]
\begin{center}
  \setlength{\unitlength}{0.91mm}
\begin{picture}(120,80)   
 \put(0,0){
  \centerline{\epsfxsize=100\unitlength
   \epsfbox{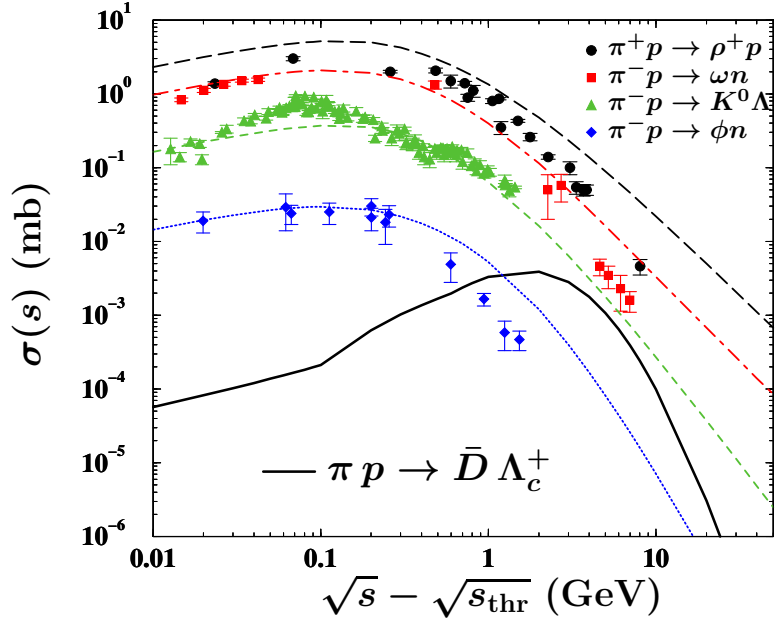}}}
 \put(48,-5){\bfseries\Large
  $\boldsymbol{\sqrt{s}-\sqrt{s_\text{thr}}~\mathrm{(GeV)}}$}
 \put(3,30){\bfseries\Large
  \begin{sideways}
   $\boldsymbol{\sigma(s)~\mathrm{(mb)}}$
  \end{sideways}}
  \put(49,13.5){\sffamily\Large
    $\boldsymbol{\pi \, p \to {\bar D} \, \Lambda_c^+}$}
  \put(90,75.5){\sffamily
    $\boldsymbol{\pi^+ p \to \rho^+ p}$}
  \put(90,71.5){\sffamily
    $\boldsymbol{\pi^- p \to \omega n}$}
  \put(90,67.5){\sffamily
    $\boldsymbol{\pi^- p \to K^0 \Lambda}$}
  \put(90,63.5){\sffamily
    $\boldsymbol{\pi^- p \to \phi n}$}
\end{picture}
\end{center}
\vspace{0.2cm}
 \caption{\label{LykIllar-fig1.eps}%
The experimental cross sections for the reactions $\pi^+ p \to \rho^+ p$,
$\pi^- p \to \omega n$, $\pi^- p \to K^0 \Lambda$ and $\pi^- p \to \phi n$
from Ref.~\protect\cite{Landoldt:1988} 
and the theoretical calculations, see the text.
The lower solid line is the prediction for the process 
$\pi \, p \to {\bar D} \, \Lambda_c^+$ within the Regge theory.
}%
\end{figure}

In Fig.~\ref{LykIllar-fig1.eps} the total cross section of 
$\pi \, p \to \bar{D} \, \Lambda_c$ reactions as a function 
of $\sqrt{s}-\sqrt{s_\text{thr}}$ is presented, here $\sqrt{s_\text{thr}}$
is the threshold energy in the c.m.s. The experimental data on 
$\pi \, N \to \rho(\omega,\phi)$ and $\pi^- \, p \to K^0 \, \Lambda^0$
are presented to illustrate their enhancement at the energy close to the
threshold, similarly to the one in the process discussed. 
The theoretical calculations of the cross sections have been done within
the Regge theory including the suppression of the open ($K,\Lambda$) and
hidden ($\phi$) strangeness production in comparison to the $\rho(\omega)$
one in the inclusive $\pi \, p$ reactions at $z \to 1$, where $z$ is
the momentum fraction of the produced hadron.

\begin{figure}[t]
\begin{center}
  \setlength{\unitlength}{0.91mm}
\centerline{\epsfxsize=100\unitlength
   \epsfbox{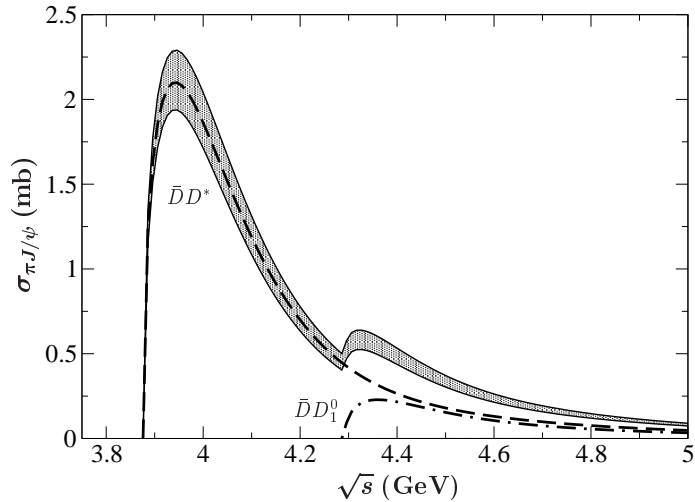}}
%
\end{center}
\vspace{0.2cm}
 \caption{\label{Fig:Jpsi-pi}%
The energy dependence of the total $\pi J/\psi$ cross sections
in the Regge model including the absorption corrections.
The figure shows also all significant partial cross sections open
to $\sqrt{s} = 5~\text{GeV}$.
The total cross section includes charge conjugation final states
where appropriate.
The estimated range of uncertainty, due to parameter variation,
is shown as a shaded band.
}%
\end{figure}

\begin{figure}[t]
\begin{center}
  \setlength{\unitlength}{0.91mm}
\centerline{\epsfxsize=100\unitlength
   \epsfbox{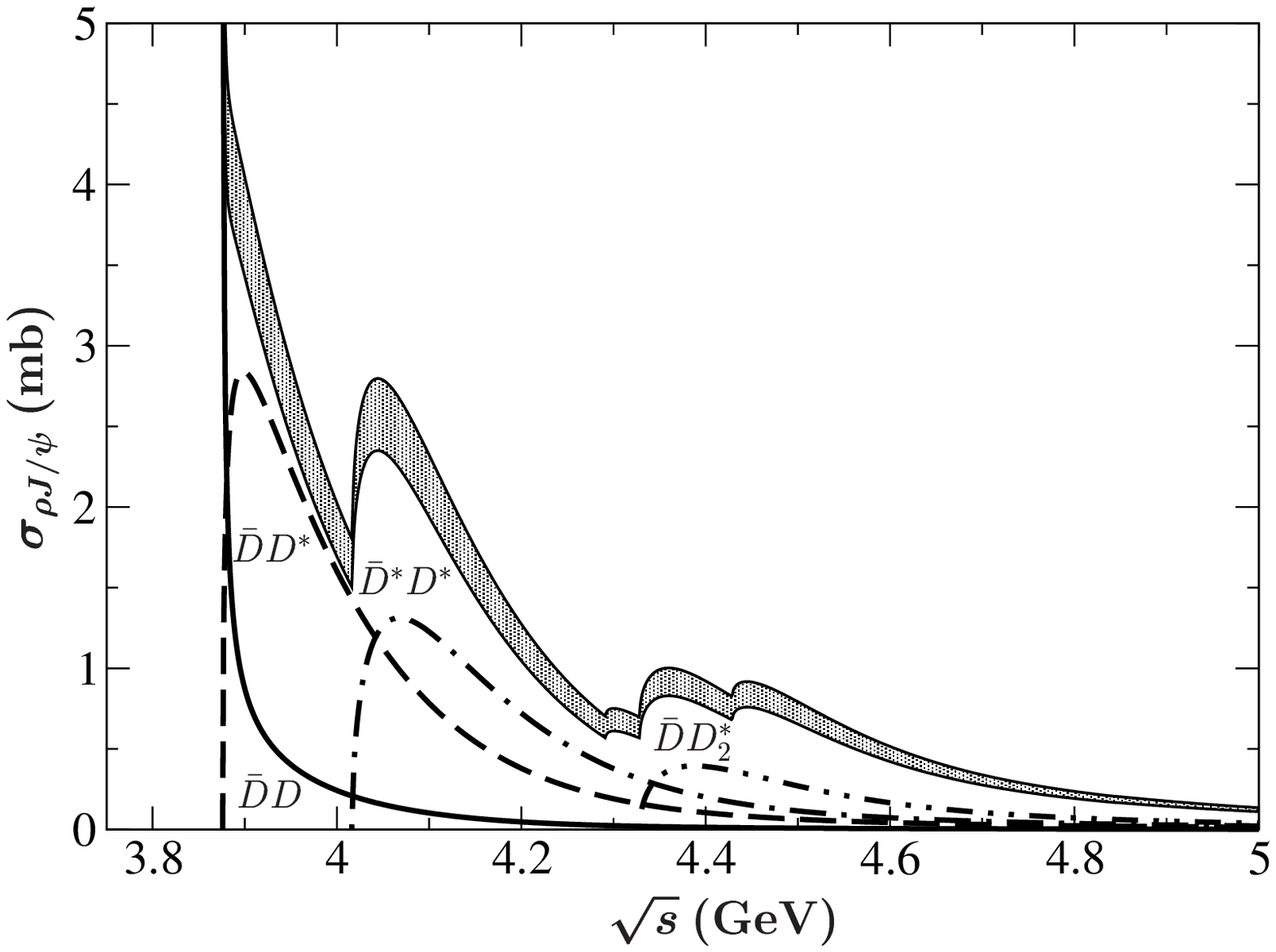}}
%
\end{center}
\vspace{0.2cm}
 \caption{\label{Fig:Jpsi-rho}%
Same as in Fig.~\ref{Fig:Jpsi-pi}, but for $\rho J/\psi$ cross sections.
}%
\end{figure}

In Figs.~\ref{Fig:Jpsi-pi},\ref{Fig:Jpsi-rho} the energy dependence of the cross
section of $D$-meson production in binary $\pi(\rho) \, J/\psi$ reactions
is presented. It is seen from Figs.~\ref{Fig:Jpsi-pi},\ref{Fig:Jpsi-rho}
that the maximum values of these cross sections at
$s > s_\text{thr}$ are a few mb. Contrary to the results obtained within
the Lagrangian model \cite{Haglin:2000,Lin:2001,Oh:2000}, all these cross
sections and the ones for $\pi \, p \to \bar{D} \, \Lambda_c(\Sigma_c)$
processes decrease when $s$ increases according to the true Regge
asymptotics.

Note, that the absorption corrections decrease the magnitude of the
cross sections discussed greatly at energies close to the threshold
and can be neglected at large $s$.

\section*{Acknowledgments}
The authors are indebted to D.~Blaschke, A.~B. Kaidalov, A.~Capella,
W.~Cassing and M.~A. Ivanov for many helpful discussions.
A.Yu.I. was supported in part by the RFBR under Grant no.~03--02--16877.

\end{document}